\documentclass[conference]{IEEEtran}
\IEEEoverridecommandlockouts

\usepackage{cite}
\usepackage{amsmath,amssymb,amsfonts}
\usepackage{algorithmic}
\usepackage{graphicx}
\usepackage{float}
\usepackage{comment}
\usepackage{subcaption}
\usepackage{multirow}
\usepackage{textcomp}
\usepackage{makecell}
\usepackage{xcolor}
\def\BibTeX{{\rm B\kern-.05em{\sc i\kern-.025em b}\kern-.08em
    T\kern-.1667em\lower.7ex\hbox{E}\kern-.125emX}}
\begin{document}

\title{Debate-Driven Multi-Agent LLMs for Phishing Email Detection\\}

\author{\IEEEauthorblockN{Ngoc Tuong Vy Nguyen}
\IEEEauthorblockA{\textit{Department of Computer Science} \\
\textit{Earlham College}\\
Richmond, IN, USA \\
nnguyen24@earlham.edu}
\and
\IEEEauthorblockN{Felix D Childress}
\IEEEauthorblockA{\textit{Department of Computer Science} \\
\textit{Earlham College}\\
Richmond, IN, USA \\
fdchild22@earlham.edu}
\and
\IEEEauthorblockN{Yunting Yin}
\IEEEauthorblockA{\textit{Department of Computer Science} \\
\textit{Earlham College}\\
Richmond, IN, USA \\
yinyu@earlham.edu}
}

\maketitle

\begin{abstract}
Phishing attacks remain a critical cybersecurity threat. Attackers constantly refine their methods, making phishing emails harder to detect. Traditional detection methods, including rule-based systems and supervised machine learning models, either rely on predefined patterns like blacklists, which can be bypassed with slight modifications, or require large datasets for training and still can generate false positives and false negatives. In this work, we propose a multi-agent large language model (LLM) prompting technique that simulates debates among agents to detect whether the content presented on an email is phishing. Our approach uses two LLM agents to present arguments for or against the classification task, with a judge agent adjudicating the final verdict based on the quality of reasoning provided. This debate mechanism enables the models to critically analyze contextual cue and deceptive patterns in text, which leads to improved classification accuracy. The proposed framework is evaluated on multiple phishing email datasets and demonstrate that mixed-agent configurations consistently outperform homogeneous configurations. Results also show that the debate structure itself is sufficient to yield accurate decisions without extra prompting strategies.

\end{abstract}


\section{Introduction}
Phishing attacks are one of the most prevalent and damaging cybersecurity threats. Attackers use feelings of fear and urgency to manipulate victims and deceive them into revealing sensitive credentials and financial information. According to industry reports, phishing remains a dominant vector for cybercrime, with a growing variety in attack strategies. Despite advancements in detection mechanisms, adversaries continuously evolve their tactics, using social engineering and obfuscation techniques to bypass automated filters. Traditional phishing detection methods primarily rely on rule-based filtering. While they are effective for detecting known phishing patterns, they often fail against novel ones. Moreover, machine learning models require extensive labeled datasets and can struggle with generalization when exposed to new forms of phishing. More recently, large language models (LLMs) have demonstrated exceptional text understanding capabilities, making them viable candidates for phishing detection. Unlike traditional classifiers, LLMs can analyze linguistic details, and even context and intent in emails given their extensive training on large datasets, allowing them to detect phishing emails intended for psychological manipulation.

\begin{figure}[H]
\centering
  \includegraphics[width=1.0\linewidth]{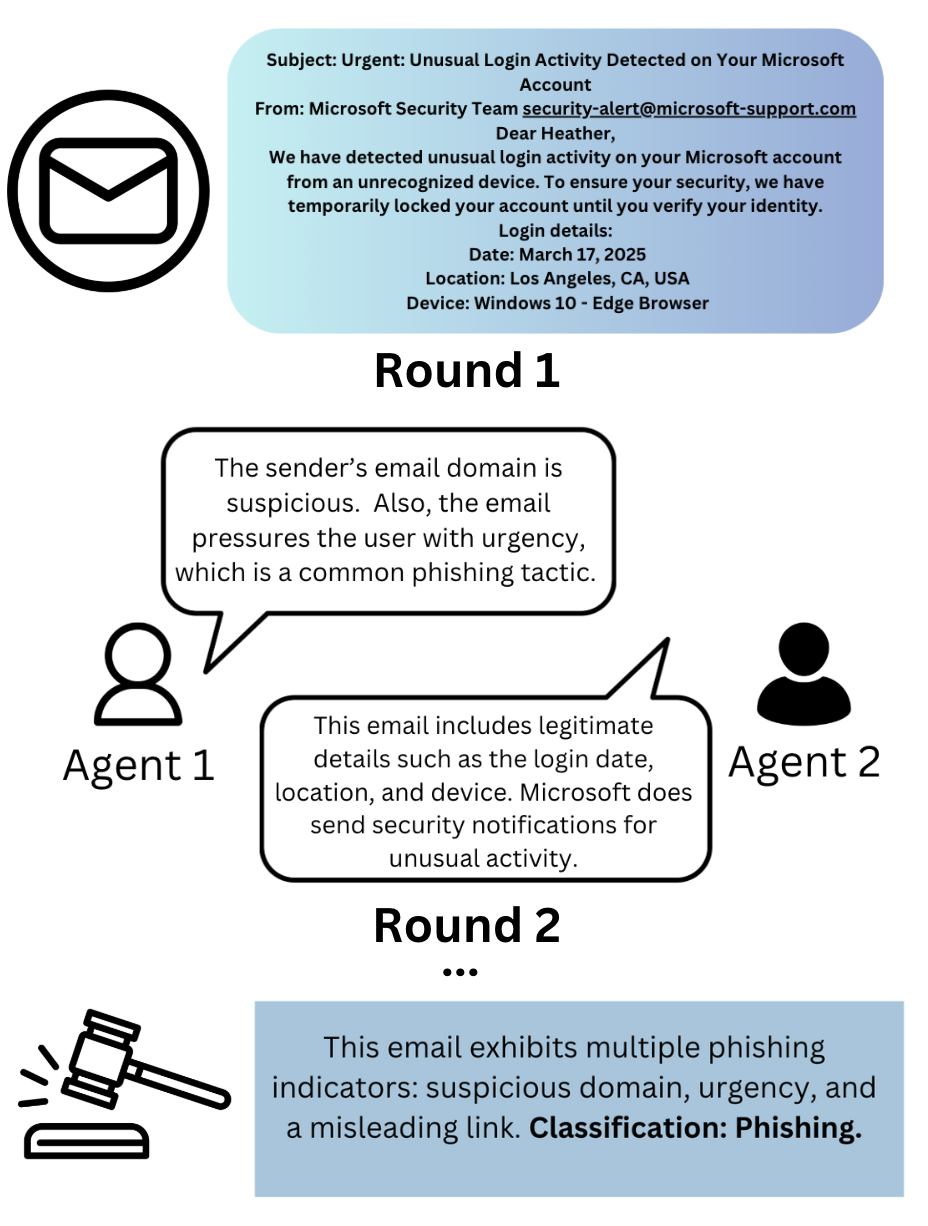}
\caption{An illustration of the proposed multi-agent debate framework, where two LLM agents engage in a structured debate over a potential phishing email, and a judge agent evaluates their arguments to produce a final classification.}
\label{fig:debate_procedure}
\end{figure}

However, a major limitation of single-agent LLM approaches is confirmation bias, where the model tends to overfit its initial reasoning path and fails to explore alternative interpretations. Additionally, phishing detection often requires contextual analysis from different perspectives. For example, an email might appear legitimate at first glance, but deeper analysis of inconsistencies or urgency cues may indicate possibility of phishing. To address these challenges, we propose a debate-driven multi-agent LLM framework for phishing email detection. Our approach employs multiple LLMs that engage in structured debates to determine whether a given message is phishing or ham (legitimate). Each agent assumes a specific stance: one arguing that the message is fraudulent and another arguing for its legitimacy. After two rounds of argument exchange, a judge agent evaluates the debate and issues a final classification decision based on the strength of the presented arguments.

Our results show that debater agent pairs with heterogeneous models outperform homogeneous ones in phishing email classification accuracy. Furthermore, we evaluate the impact of prompt engineering techniques including chain-of-thought and role prompting, and find that they do not significantly improve performance on top of our debate framework. These findings prove the value of multi-agent reasoning in phishing detection and suggest broader potential for debate-based frameworks in other NLP and cybersecurity applications.

This paper is organized as follows: In Section \ref{sec:relatedworks}, we review prior literature on phishing detection techniques and the use of multi-agent LLM systems. Section \ref{sec:datasets} describes the datasets used in our experiments and the preprocessing steps taken to ensure compatibility with LLM input constraints. Section \ref{sec:methods} describes our proposed multi-agent debate framework, including the design of prompt templates and the debate procedure. In Section \ref{sec:experiments}, we present experimental results comparing different agent configurations and prompting strategies across multiple datasets. Finally, we conclude our findings in Section\ref{sec:conclusion}.

\section{Related Works}
\label{sec:relatedworks}
In this section, we present a comprehensive literature review on phishing attacks, their impact, and the various approaches used for detection and mitigation. We survey how machine learning, deep learning, and LLM-based techniques have been used to combat phishing threats and analyze their effectiveness and limitations. Additionally, we explore the use of multi-agent debate systems in various computing tasks before proposing such a framework for phishing detection.

\subsection{Phishing Email Attack Detection}
Phishing email attacks are a form of cyberattack where attackers disguise themselves as trustworthy individuals or organizations to deceive recipients of their emails into taking harmful actions. These emails often attempt to steal sensitive information such as passwords, credit card details, or personal data \cite{b1}. They also attempt to spread malware and manipulate users into making financial transactions. Phishing emails typically include links that redirect users to malicious websites or attachments embedded with malware. Attackers often employ social engineering tactics, such as creating a sense of urgency or fear, or impersonating authority figures to trick victims into responding. Phishing emails can be difficult for humans to detect, especially when they look similar to safe emails \cite{b2}. Phishing email detection is a well-studied area in cybersecurity, and various techniques have been developed over time, ranging from traditional methods to machine learning and deep learning approaches. 

Traditional detection methods include blacklist \cite{b3} and whitelist\cite{b4, b5} techniques that block emails from known malicious sources or only allow emails from trusted senders. However, phishing email attacks are evolving to bypass these traditional detection methods by closely mimicking legitimate communications and using sophisticated evasion techniques, such as dynamic URLs and AI-generated content \cite{b6}. As a result, cybersecurity defenses have transitioned from simple rule-based filters to advanced machine learning and deep learning models that derive patterns from email structure, language, and metadata to improve detection accuracy. Salahdine et al. \cite{b7} propose a machine learning-based phishing detection technique by extracting 10 key features from 4,000 phishing emails. Experimental results on classification task demonstrate that an artificial neural network achieves superior accuracy compared to other methods. Valecha et al. \cite{b8} investigate the effectiveness of persuasion cues, specifically gain and loss cues, in phishing email detection by developing three machine learning models that incorporate these cues. The results show that persuasion cues improve model performance, and that psychological tactics are useful in anti-phishing methods. Hamid and Abawajy \cite{b9} propose a technique to improve the accuracy of phishing email detection by using a hybrid feature selection method combining content-based and behavior-based features extracted from email headers. The proposed method has improved detection rates due to successful identification of attacker behaviors. Altwaijry et al. \cite{b10} explore one-dimensional CNN-based models with integrated recurrent layers for phishing email detection, and show that the 1D-CNNPD model with Bi-GRU achieves the highest accuracy and demonstrates the potential of deep learning techniques to reduce false positive rate.

With the increasing popularity of LLMs, they are being increasingly used for developing phishing detection tools. Koide et al. \cite{b11} propose a phishing email detection system called ChatSpamDetector that uses GPT-4 to analyze email content and provide both classification and explanatory reasoning. Experimental results demonstrate that ChatSpamDetector outperforms traditional spam filters and baseline machine learning models by effectively identifying phishing tactics. Heiding et al. \cite{b12} compare phishing emails generated by GPT-4, V-Triad, and a combination of both, and conclude that LLM-generated content generally outperform generic phishing emails. Thy also argue that LLMs can effectively detect phishing intent and sometimes outperform human detection performance. Lee \cite{b13} uses hybrid feature selection and prompt engineering to investigate the effectiveness of LLMs in detecting phishing emails of various types, including spear phishing, traditional phishing, and AI-generated phishing. Experimental results show that Llama-3.1-70B achieves superior accuracy over other models while also provides interpretable reasoning.

\subsection{Multi-agent Debate to Enhance LLM Reasoning}

Recent research has showed that multi-agent debate frameworks are useful techniques to enhance the reasoning abilities of LLMs. Unlike single-agent prompting methods, multi agent setups create multiple LLM instances to critique and refine each other’s responses, which often lead to more accurate and well-reasoned outputs. Du et al. \cite{b14} introduce a multi-agent debate approach where multiple LLM instances debate their responses over multiple rounds to improve reasoning, and their findings show that this method improves mathematical and strategic reasoning while reducing hallucinations. Liang et al. \cite{b15} identify the Degeneration-of-Thought problem in LLMs, where self-reflection fails to generate novel insights once the model becomes overconfident in its initial solution. They address the identified problem with a multi-agent debate framework with judge supervision, and concluded that the reasoning performance is improved on complex tasks like commonsense translation and arithmetic. Further investigating the use of multi-agent interactions, Estornell et al. \cite{b16} propose ACC-Collab, a learning framework that trains a two-agent team that contains an actor-agent and a critic-agent to facilitate collaboration between the agens and improve their problem-solving skills. The proposed framework outperforms existing multi-agent methods in various benchmarks. Wang et al. \cite{b17} make two observations on multi-agent discussions: they outperform single-agent setups when no demonstrations are provided, and in multi-LLM environments, stronger LLMs help weaker ones reason through interaction. Collectively, these studies demonstrate the effectiveness of multi-agent debate frameworks in improving decision-making and reasoning abilities of LLMs. They also demonstrate the potential of such frameworks in cybersecurity applications, such as phishing email detection studied in this work, where reasoning over text is critical.

\section{Datasets}
\label{sec:datasets}
To evaluate the effectiveness of our proposed multi-agent debate framework, we use a diverse set of email datasets that capture different communication styles and phishing tactics. These datasets include both real-world and synthetically generated emails, spanning multiple time periods. Incorporation of multiple sources makes our experimental setup representative of the diverse scenarios typically encountered in phishing email detection. The datasets used in our study are as follows:
\begin{itemize}
    \item \textbf{University of Twente Phishing Validation Emails Dataset \cite{b18}:} A dataset of 2,000 real and artificially generated example emails. 
    \item \textbf{Eleven Curated Datasets of Phishing Email \cite{b19, b20}:} A collection of cleaned email corpus with phishing or ham labels, including CEAS-08, Ling-Spam, Enron, Nazario, Nazario\_5, Nigerian\_5, Nigerian\_Fraud, SpamAssassin, TREC-05, TREC-06, and TREC-07.
\end{itemize}

To keep the number of input tokens within the practical limits of LLMs, we selected the University of Twente (UoT) dataset along with a representative subset of four curated datasets from the publicly available corpus collection. Specifically, Ling, Nazario\_5, Nigerian\_Fraud, and SpamAssasin are used for our experiments. The summary statistics of these datasets are presented in Table \ref{tab:dataset_summary}. During exploratory data analysis, we observed that certain datasets contained emails with exceptionally long content, which could exceed token limits imposed by LLM architectures. To address this, we kept only the emails whose token lengths fall within the 75th percentile of their respective dataset distributions. After applying this filtering step, the final dataset used in our experiments comprises a total of 12,798 emails, of which 6,475 are ham (safe) and 6,323 are phishing.

\begin{table}[h]
    \centering
    \begin{tabular}{|c|c|cc|c|c|}
        \hline
        \multirow{2}{*}{\textbf{Dataset}} & \multirow{2}{*}{\textbf{Size}} & \multicolumn{2}{c|}{\textbf{Email Length}} & \multirow{2}{*}{\makecell{\textbf{Num of} \\ \textbf{Ham}}} & \multirow{2}{*}{\makecell{\textbf{Num of} \\ \textbf{Phishing}}} \\
        \cline{3-4}
        &  & \textbf{Avg} & \textbf{75\%} &  &  \\
        \hline
        UoT  & 2000  & 86.71  & 95.00  & 1000  & 1000  \\
        Ling  & 2859 & 3222.32 & 4014.50 & 2401 & 458 \\
        Nazario\_5  & 3065 & 3545.33 & 1630.00 & 1500 & 1565 \\
        Nigerian\_Fraud  & 3332 & 2644.38 & 3211.75 & 0 & 3332 \\
        SpamAssasin  & 5808 & 2406.91 & 2028.00 & 4091 & 1718 \\
        \hline
    \end{tabular}
    \caption{Summary statistics of selected phishing and ham email datasets, including size, email length distribution, and label distribution.}
    \label{tab:dataset_summary}
\end{table}

\begin{table*}[ht]
\centering
\caption{Phishing email detection accuracy and F1 scores on five benchmark datasets, comparing different agent configurations and the effects of prompt engineering techniques.}
\label{tab:combined_table}
\begin{tabular}{|c|c|c|cc|cc|cc|cc|cc|}
\hline
\multicolumn{3}{|c|}{\textbf{Agent Configuration}} &
\multicolumn{2}{c|}{\textbf{UoT}} &
\multicolumn{2}{c|}{\textbf{Ling}} &
\multicolumn{2}{c|}{\textbf{Nazario\_5}} &
\multicolumn{2}{c|}{\textbf{Nigerian\_Fraud}} &
\multicolumn{2}{c|}{\textbf{SpamAssassin}} \\
\cline{1-3} \cline{4-13}
\textbf{Agent 1} & \textbf{Agent 2} & \textbf{Judge} & \textbf{Acc} & \textbf{F1} & \textbf{Acc} & \textbf{F1} & \textbf{Acc} & \textbf{F1} & \textbf{Acc} & \textbf{F1} & \textbf{Acc} & \textbf{F1} \\
\hline
GPT-4 & GPT-4 & GPT-4 & 98.12\% & 0.98 & 98.76\% & 0.98 & 98.03\% & 0.98 & 98.54\% & / & 98.40\% & 0.98 \\
LLaMA-2 & LLaMA-2 & LLaMA-2 & 98.01\% & 0.98 & 98.32\% & 0.98 & 98.22\% & 0.98 & 98.17\% & / & 98.09\% & 0.98 \\
GPT-4 & LLaMA-2 & GPT-4 & \textbf{98.91\%} & \textbf{0.98} & \textbf{99.43\%} & \textbf{0.99} & \textbf{99.02\%} & \textbf{0.99} & \textbf{99.27\%} & / & 98.73\% & 0.98 \\
LLaMA-2 & GPT-4 & GPT-4 & 98.36\% & 0.98 & 99.02\% & 0.99 & 98.71\% & 0.98 & 98.85\% & / & \textbf{99.12\%} & \textbf{0.99} \\
\hline
\multicolumn{3}{|c|}{GPT-4–LLaMA-2–GPT-4-CoT} & 98.65\% & 0.98 & 99.12\% & 0.99 & 98.77\% & 0.98 & 99.00\% & / & 98.63\% & 0.98 \\
\multicolumn{3}{|c|}{GPT-4–LLaMA-2–GPT-4-Role} & 98.65\% & 0.98 & 98.95\% & 0.98 & 98.52\% & 0.98 & 98.74\% & / & 98.90\% & 0.99 \\
\multicolumn{3}{|c|}{GPT-4–LLaMA-2–GPT-4-CoT-Role} & 98.38\% & 0.98 & 98.77\% & 0.98 & 98.69\% & 0.98 & 98.59\% & / & 98.45\% & 0.98 \\
\hline
\end{tabular}
\label{tab:accuracy_results}
\end{table*}

\section{Methods}
\label{sec:methods}
\subsection{Multi-Agent Debate Framework}
\label{sec:Multi-Agent_Debate_Framework}
We propose a multi-agent debate framework for phishing email detection, composed of three components: two debater agents, a pre-defined and scripted debate procedure, and a judge agent. The debater agents consist of two LLM-based instances, which may be instantiated from the same or different models. The first agent is prompted to argue that the given email is a phishing attempt, while the second agent is prompted to respond to the first agent's output by countering those claims and arguing for the email’s legitimacy. The two agents then engage in another round to make sure that the arguments are well-formulated while maintaining computational efficiency. 

The debate procedure is pre-defined and scripted to generate template prompts for each email in the dataset:

\begin{enumerate}
    \item Round One: 
    \begin{itemize}
        \item Carefully analyze the following email and argue why it is likely to be a phishing attempt \textbf{(Agent 1)}
        \item Carefully analyze the following email and argue why it is likely to be legitimate and not a phishing attempt \textbf{(Agent 2)}
    \end{itemize}

    \item Round Two: 
        \begin{itemize}
            \item Given your opponent's rebuttal, reinforce your position that the following email is a phishing attempt \textbf{(Agent 1)}
            \item Given your opponent's rebuttal, reinforce your position that the following email is not a phishing attempt \textbf{(Agent 2)}
        \end{itemize}
\end{enumerate}

Arguments made by the two agents are logged for subsequent judge evaluation. Following the two-round debate, a third LLM instance serving as the judge agent is fed the four arguments and is prompted to evaluate the strength and coherence of the arguments. The judge is then prompted to provide a final binary classification verdict, phishing or legitimate, which is logged alongside its reasoning for performance assessment.

\subsection{Synergy with Prompt Engineering Techniques}
We also incorporate two prompting engineering techniques in our multi-agent debate framework to evaluate their effectiveness on phishing email detection.

\subsubsection{Chain-of-Thought Prompting}
Chain-of-Thought (CoT) prompting is a technique that encourages language models to generate intermediate reasoning steps before arriving at a final answer. In order to guide the agents to articulate the rationale behind their judgment, the following prompts are appended to the end of the basic template shown in Section \ref{sec:Multi-Agent_Debate_Framework}:

\begin{itemize}
    \item \textbf{Agent 1:} Break down your reasoning \textbf{step-by-step} using these guiding questions:
        1. Is the language designed to invoke urgency, fear, or greed?
        2. Are there misleading links or unusual requests?
        3. Does the email resemble common phishing patterns?
    \item \textbf{Agent 2:} Break down your reasoning \textbf{step-by-step} using these guiding questions:
	1.	Is the tone and language professional and consistent?
	2.	Are the links safe and are the requests expected?
	4.	Does the context match what a legitimate sender would send?
\end{itemize}

\subsubsection{Role Prompting}
Role prompting is a prompting technique to instruct a language model to take on a specific role or persona to influence its tone and reasoning. Instead of simply asking the model to perform a task, it is first told who it is in the context of the task. This technique leads to more coherent and contextually appropriate responses, and also prevents agents from echoing each other by further anchoring them in distinct roles. To assign roles for the agents, the following prompts are appended to the front of the basic template:
\begin{itemize}
    \item \textbf{Agent 1:} You are a senior cybersecurity analyst at a large tech company. Your job is to review suspicious emails reported by employees and determine that they are phishing attempts.
    \item \textbf{Agent 2:} You are an email forensics expert working for an IT compliance team. Your job is to validate that a flagged email is legitimate and not a phishing attempt.
\end{itemize}

\section{Experiments}
\label{sec:experiments}
To evaluate the effectiveness of our proposed multi-agent debate framework, we conduct experiments using different combinations of two LLMs, GPT-4 and LLaMA-2, as debater agents and judges. The different agent-agent-judge configurations allow us to analyze the impact of different model pairings on debate performance and classification accuracy.

For each email in the filtered dataset described in Section \ref{sec:datasets}, the debate procedure was executed using the prompt templates detailed in Section \ref{sec:Multi-Agent_Debate_Framework}. The arguments from both agents were collected, and the judge agent was prompted to evaluate the debate and produce a binary classification label, along with a brief justification. The predicted labels were then compared to the ground-truth labels to compute classification accuracy across the five datasets.

Table \ref{tab:accuracy_results} presents phishing email detection accuracy for each agent-agent-judge setup, evaluated on the five selected datasets: UoT, Ling, Nazario\_5, Nigerian\_Fraud, and SpamAssassin. We observe that both the fully GPT-4 and fully LLaMA-2 configurations consistently underperform relative to mixed-model setups. Notably, the mixed configuration GPT-4–LLaMA-2–GPT-4 achieves the highest accuracy on four out of the five datasets, indicating that heterogeneous agents can complement each other’s reasoning capabilities. This finding supports the observations made by Wang et al. \cite{b17}, which state that multi-agent systems often outperform single-agent setups, and that collaboration between agents can improve their task performance.

To further investigate the factors contributing to model performance within the debate framework, we conducted experiments evaluating the impact of chain-of-thought prompting, role prompting, and the combination of both, with the best-performing agent configuration (GPT-4–LLaMA-2–GPT-4) as the baseline. As shown in Table \ref{tab:accuracy_results}, neither CoT prompting, role prompting, nor their combination outperformed the baseline configuration without these enhancements. While CoT prompting encourages step-by-step reasoning and role prompting assigns each agent a distinct persona, the differences in reasoning did not translate into measurable gains in classification accuracy or F1 score across the datasets.

One possible explanation is that the debate framework itself already results in sufficiently structured reasoning, particularly with the multi-round interaction procedure, as the exchange of arguments and counterarguments naturally prompts each agent to support its position with evidence. Furthermore, the predefined prompt template already assigns opposing positions to the two agents, as one must advocate that the email is phishing while the other defending its legitimacy. As a result, explicit role prompting may offer limited added value.

These results show the effectiveness of multi-agent debate in phishing email detection and highlight the importance of selecting a mixture of capable agents for both argumentation and judgment. In particular, our findings suggest that using mixed models is critical for accurate final decisions, and that they have a greater impact than added prompting complexity.

\section{Conclusion}
\label{sec:conclusion}
In this work, we present a multi-agent large language model debate framework for phishing email detection. Unlike traditional rule-based or single-model classifiers, our approach simulates a structured argument between two LLM agents followed by a third judge agent that issues the final classification verdict. Our experiments on five benchmark phishing datasets demonstrate that mixed-model agent configurations consistently outperform homogeneous setups. These results support the hypothesis that heterogeneous agents can complement each other’s reasoning abilities, leading to more accurate classification outcomes. Additionally, we explored the impact of two prompting strategies, chain-of-thought and role prompting, but found that they did not significantly improve performance over the best baseline. This suggests that the structured debate mechanism itself already elicits rich reasoning without additional prompting engineering efforts.


\end{document}